\documentclass[12pt]{article}
\usepackage{amssymb,amsmath}
\usepackage[noblocks]{authblk}
\usepackage[top=0.75in, bottom=0.75in, left=0.75in, right=0.75in, dvips]{geometry}
\usepackage{caption}
\begin{document}
\textwidth 10.0in 
\textheight 9.0in 
\topmargin -0.60in
\title{The Non-Abelian Gauge Field Propagator in a\\
Plane Wave Background Field}
\author[1,2]{D.G.C. McKeon}
\affil[1] {Department of Applied Mathematics, The
University of Western Ontario, London, ON N6A 5B7, Canada} 
\affil[2] {Department of Mathematics and
Computer Science, Algoma University, \newline Sault Ste. Marie, ON P6A
2G4, Canada}
\date{}
\maketitle         

\maketitle         
\noindent
email: dgmckeo2@uwo.ca\\
PACS No.: \\
KEY WORDS:

\begin{abstract}
Employing methods introduced by Schwinger in quantum electrodynamics, we compute the propagator for a non-Abelian gauge field in a plane wave background field. In the long distance limit a mass-like term for the gauge field is induced by this interaction.
\end{abstract}

\section{Introduction}
The one-loop effective action arising due to the interaction of an electron with a constant background electromagnetic field was examined by Euler and Heisenberg [1] with a charged scalar field treated by Weisskopf [2].  Schwinger also treated this problem and in addition treated a plane wave background field [3].  Subsequently, there has been a large effort to examine how these calculations could be extended; for reviews see refs. [4,5,6].

The techniques employed in quantum electrodynamics can also be applied to external fields in models involving non-Abelian gauge fields [7,8].

Although a background plane wave field cannot contribute to the effective action in quantum electrodynamics [3,5], it does contribute to the propagator [3,4], in part by providing an effective mass to the electron.  We shall show that in non-Abelian gauge theory a similar effect occurs; a background plane wave field can provide an effective mass to the vector.  Since interaction with a background field can be viewed as emission of low energy quanta, this may affect the infrared behaviour of the theory.

\section{A Background Gauge Field}
The Lagrangian for a non-Abelian vector field $V_\mu^a$
\begin{equation}
\mathcal{L} = - \frac{1}{4} F_{\mu\nu}^a (V) F^{a\mu\nu}(V)
\end{equation}
when $V_\mu^a$ is expanded in terms of a background field $A_\mu^a$ and a quantum field $Q_\mu^a \left(V_\mu^a = A_\mu^a + Q_\mu^a\right)$ has a contribution bilinear in $Q_\mu^a$
\begin{equation}
\mathcal{L}^{(2)} = \frac{1}{2}Q_\mu^a \left[ \eta^{\mu\nu}(D_\lambda(A)D^\lambda(A))^{ab} + 2g\epsilon^{apb}F^{p\mu\nu}(A)\right]Q_\nu^b.
\end{equation}
We use the notation $F_{\mu\nu}^a(V) = \partial_\mu V_\nu^a - \partial_\nu V_\mu^a + g\epsilon^{abc}V_\mu^bV_\nu^c$ and $D_\mu^{ab}(V) = \partial_\mu \delta^{ab}+ g\epsilon^{apb}V_\mu^p$ as well as the gauge fixing 
\begin{equation}
\mathcal{L}_{gf} = -\frac{1}{2} \left(D_\mu^{ab} (A)Q^{b\mu}\right)^2.
\end{equation}
The background field that we use is
\begin{equation}
A_\mu^a (x) = \epsilon_\mu^z f(\xi)
\end{equation}
where $\epsilon_\mu^z$ is a vector oriented in the particular direction ``$z$'' in group space and $\xi = n \cdot x$ with
\begin{equation}
n \cdot n = n \cdot \epsilon^z = 0, \quad \epsilon^z \cdot \epsilon^z = -1.
\end{equation}
We see that the propagator for the quantum field $Q$ involves [3]
\begin{equation}
M_{xx^\prime} = <x| e^{-iH\tau}|x^\prime>
\end{equation}
where 
\begin{equation}
H_{\mu\nu}^{ab}= \eta_{\mu\nu}\left(\pi_\lambda \pi^\lambda\right)^{ab} - 2g\epsilon^{apb} F_{\mu\nu}^p
\end{equation}
and
\begin{equation}
\left(\pi_\lambda\right)_{\mu\nu}^{ab} = iD_\lambda^{ab} \eta_{\mu\nu} = \left( p_\lambda \delta^{ab} + i\epsilon^{apb} A_\lambda^p (x)\right)\eta_{\mu\nu}.
\end{equation}
Since
\begin{subequations}
\begin{align}
\left[ x_\mu , p_\nu \right] &= -i\eta_{\mu\nu}\\
\left[ \pi_\mu , \pi_\nu \right]^{ab} &= -g\epsilon^{apb} F_{\mu\nu}^p
\end{align}
\end{subequations}
it follows that
\begin{subequations}
\begin{align}
\left( \frac{dx_\lambda}{d\tau}\right)_{\mu\nu}^{ab} &= i \left[H, x_\lambda\right]^{ab}_{\mu\nu} = -2\left(\pi_\lambda\right)_{\mu\nu}^{ab}\\
\intertext{and}
\left( \frac{d\pi_\lambda}{d\tau}\right)_{\mu\nu}^{ab} &= i \left[H, \pi_\lambda\right]^{ab}_{\mu\nu} = \left[ -2ig \epsilon^{apq}F_{\rho\lambda}^p \pi^{\rho qb} + g\epsilon^{apb} \left(D^{\rho pq}F_{\rho\lambda}^q\right)\right]\eta_{\mu\nu}\\
&\hspace{2cm} -2g\left( D_\lambda^{ap} \epsilon^{pqb}F_{\mu\nu}^q\right).\nonumber
\end{align}
\end{subequations}
With the background field of eq. (4), we have
\begin{equation}
F_{\mu\nu}^a = \left( n_\mu \epsilon_\nu^z - n_\nu \epsilon_\mu^z\right) f^\prime (\xi) \equiv \phi_{\mu\nu}^z f^\prime (\xi)
\end{equation}
and so by eqs. (10b)
\begin{equation}
\left( \frac{d\pi_\lambda}{d\tau} \right)^{ab}_{\mu\nu} = \left( -2ig\epsilon^{azp} \phi_{\rho \lambda}^z f^\prime(\xi) \pi^{\rho ab} \right) \eta_{\mu\nu} -2g\epsilon^{azb} n_\lambda \phi_{\mu\nu}^z f^{\prime\prime}(\xi)\;.
\end{equation}
Eqs. (5, 10a, 12) now lead to 
\begin{subequations}
\begin{align}
\frac{d\;n\cdot \pi}{d\tau} &= 0\\
\intertext{and}
\frac{d\xi}{d\tau} &= -2n\cdot \pi ,
\end{align}
\end{subequations}
and so $n \cdot \pi$ is a constant and
\begin{equation}
\xi (\tau) = \xi(0) -2 n\cdot \pi \tau .
\end{equation}
Furthermore, we find by eqs. (5, 12) that
\begin{equation}
\epsilon^{azp} \phi_{\rho\sigma}^z \pi^{\rho p b} = ig E^{ab} n_\sigma f(\xi) + K_\sigma^{ab}
\end{equation}
where $K_{\sigma}^{ab}$ is a constant, $n^\sigma K_\sigma^{ab} = 0$ and $E^{ab} \equiv 
\epsilon^{azp} \epsilon^{pzb}$. 

Using eq. (15), eq. (12) leads to
\begin{equation}
\pi_\lambda = \frac{-1}{2n\cdot \pi} \left[ g^2 E^{ab} n_\lambda f^2 - 2igK_\lambda^{ab} f-2g\epsilon^{azb} n_\lambda \phi_{\mu\nu}^z f^\prime \right] + D_\lambda^{ab}
\end{equation}
where $D_\lambda^{ab}$ is a constant.  Integration of eq. (10a) then gives us
\begin{align}
-\frac{1}{2}\left(x_\lambda(\tau) - x_\lambda(0) \right) &= \left( \frac{-1}{2n\cdot \pi}\right)^2 \int_{\xi(0)}^{\xi(\tau)} d\xi \Big[ g^2 E^{ab} n_\lambda f^2 - 2igK_\lambda^{ab} f\nonumber \\
& - 2g\epsilon^{azb} \phi_{\mu\nu}^z n_\lambda f^\prime \Big] + D_\lambda^{ab} \tau.
\end{align}
From eq. (15) it follows that 
\begin{equation}
\phi_\rho^{z\sigma} K_\sigma^{ab} = \frac{n_\rho \epsilon^{azb}(\xi(\tau)-\xi(0)}{2\tau}
\end{equation}
upon using eq. (14) and $\phi_\mu^{z\lambda}\phi_{\nu\lambda}^z = -n_\mu n_\nu$.

Using eq. (16) to eliminate $\pi_\lambda^{ab}$ from eq. (15) we obtain
\begin{align}
K_\sigma^{ab} = \epsilon^{azp} \phi^{z\lambda}_{\;\;\;\;\;\sigma} \Big[ \frac{-1}{2n\cdot\pi} & \left( g^2 E^{pb} n_\lambda f^2 - 2ig K_\lambda^{pb} f- 2g\epsilon^{pzb} n_\lambda\phi_{\mu\nu}^z f^\prime \right)\nonumber \\
& + D_\lambda^{pb} \Big] -igE^{ab} n_\sigma f;
\end{align}
as $n^\lambda \phi_{\lambda\rho}^z = 0$ eqs. (19) and (17) result in 
\begin{align}
D_\lambda^{ab} =  \frac{-1}{2\tau} \left( x_\lambda(\tau) - x_\lambda(0)\right) - \frac{\tau}{(\xi(\tau)-\xi(0))^2} \int_{\xi(0)}^{\xi(\tau)} d\xi
 & \Big[ g^2 E^{ab} n_\lambda f^2 - 2igK_\lambda^{ab} f \\
& - 2g\epsilon^{azb} n_\lambda \phi_{\mu\nu}^z f^\prime \Big].\nonumber
\end{align}
If now eq. (20) is used to eliminate $D_\lambda^{pb}$ in eq. (19) and eq. (18) is also used,
\begin{equation}
K_\sigma^{ab} = \epsilon^{azb} \phi^{z\;\;\;\rho}_{\;\;\sigma} \left( \frac{x_\rho(\tau) - x_\rho(0)}{2\tau} \right) - \frac{ig n_\sigma}{\xi(\tau) - \xi(0)} \int_{\xi(0)}^{\xi(\tau)}d\xi f(\xi) E^{ab}.
\end{equation}
After eliminating $K_\sigma^{ab}$ in eq. (20) by use of eq. (21), and then with the resulting expression for $D_\sigma^{ab}$, we find that eq. (16) becomes
\begin{align}
\pi_\sigma = \left( \frac{\xi(\tau) - \xi(0)}{\tau}\right)^{-1}  & \left( g^2 E^{ab} n_\lambda f^2 - 2ig K_\lambda^{ab} f - 2g\epsilon^{azb} n_\lambda \phi_{\mu\nu}^z f^\prime \right) \\
& + \Bigg[ - \frac{x_\lambda(\tau) - x_\lambda(0)}{2\tau} - 
\frac{\tau}{(\xi(\tau) - \xi(0))^2} \int_{\xi(0)}^{\xi(\tau)}d\xi \big( g^2 E^{ab} n_\lambda f^2 \nonumber \\
&\hspace{1cm}- 2igK_\lambda^{ab} f - 2g \epsilon^{azb} \phi_{\mu\nu}^z n_\lambda f^\prime\big) \Bigg]\nonumber
\end{align}
From eq. (21) we see that
\begin{equation}
\left( x_\lambda(\tau) - x_\lambda(0)\right) K^{ab\lambda} = -ig E^{ab}
\int_{\xi(0)}^{\xi(\tau)}d\xi f(\xi)
\end{equation}
\begin{equation}
K_\lambda^{ap} K^{pb\lambda} = -E^{ab} \left( \frac{\xi(\tau)-\xi(0)}{2\tau} \right)^2.
\end{equation}
Eqs. (22-24) along with eq. (11) reduce the Hamiltonian of eq. (7) to
\begin{align}
H &= \frac{x_\lambda^2(\tau) - 2x_\lambda (\tau)x^\lambda(0) + x_\lambda^2(0)}{(2\tau)^2} - \frac{2i}{\tau}+ \frac{2g}{\xi(\tau) - \xi(0)}\epsilon^{azb}\phi_{\mu\nu}^z 
\left( f(\xi(\tau)) - f(\xi(0))\right) \\
& + g^2 E^{ab} \left[ \frac{1}{\xi(\tau) - \xi(0)}
\int_{\xi(0)}^{\xi(\tau)}d\xi f^2(\xi)  - \left( \frac{1}{\xi(\tau) - \xi(0)} 
\int_{\xi(0)}^{\xi(\tau)}d\xi f(\xi)\right)^2\right]\nonumber
\end{align}
as by eqs. (7, 14)
\begin{equation}
\left[ \xi(0), x_\lambda(\tau) \right] = \left[ \xi(0) + 2n \cdot \pi \tau, x_\lambda (\tau)\right] = 2in_\lambda \tau
\end{equation}
\begin{equation}
\left[ x_\lambda (0), x^\lambda(\tau) \right] = 8i\tau.
\end{equation}

Since [3]
\begin{equation}
i \partial_\tau <x (\tau)|x^\prime(0) > = < x(\tau) |H| x^\prime(0) >
\end{equation}
the Hamiltonian of eq. (25) leads to
\begin{align}
<x (\tau)|x^\prime(0) >  &= \frac{C(x,x^\prime)}{\tau^2} \exp i \Bigg\{
\frac{(x-x^\prime)^2}{4\tau} - \tau \Bigg[ \frac{2g}{\xi - \xi^\prime} \epsilon^{azb} \phi_{\mu\nu}^z \left( f(\xi) - f(\xi^\prime)\right)\\
& + g^2 E^{ab} \left( \frac{1}{\xi-\xi^\prime} \int_{\xi^\prime}^\xi dyf^2(y) - \left(\frac{1}{\xi - \xi^\prime} \int_{\xi^\prime}^\xi dyf(y)\right)^2\right) \Bigg]\Bigg\}.\nonumber
\end{align}
We also have [3] from eq. (8)
\begin{equation}
iD_\lambda^{ab} <x (\tau)|x^\prime(0) > = <x (\tau)| \pi_\lambda (\tau) |x^\prime(0) >
\end{equation}
which by eq. (29) leads to 
\begin{equation}
C(x,x^\prime) = \frac{-i}{(4\pi)^2} I\!\!P \exp \left( - \int_{x^\prime}^x dy^\lambda \epsilon^{azb} A_\lambda^z (y) \right)
\end{equation}
where $I\!\!P$ denotes path ordering, upon following the procedure of ref. [3].

If now $f(y)$ in eq. (4) is identified with a sinusoidal function,
\begin{equation}
f(y) = \kappa \cos (y)
\end{equation}
where $\kappa$ is a dimensionful parameter, then
\begin{align}
\frac{1}{\xi - \xi^\prime} \int_{\xi^\prime}^\xi dyf^2(y) & - \left( \frac{1}{\xi - \xi^\prime} \int_{\xi^\prime}^\xi dy f (y)\right)^2\nonumber \\
&= \kappa^2 \left[ \frac{1}{2} + \frac{\sin(2\xi)-\sin(2\xi^\prime)}{4(\xi - \xi^\prime)} - \left( \frac{\sin\xi -\sin\xi^\prime}{\xi - \xi^\prime}\right)^2\right].
\end{align}
This term acts as an effective mass parameter in the matrix element of eq. (29).  If $\xi - \xi^\prime \rightarrow 0$ then it approaches zero, while if $\xi - \xi^\prime \rightarrow \infty$ it approaches $\kappa^2/2$, providing an effective mass for those components of $Q_\mu^a$ for which $E^{ab}$ is non zero.

\section{Discussion}
The plane wave background field of eqs. (4,11) has the property that $F_{\mu\nu}^a F^{a\mu\nu} = \frac{1}{2} \epsilon^{\mu\nu\lambda\sigma} F_{\mu\nu}^a F_{\lambda\sigma}^a = 0$ and so the effective action cannot depend on $F_{\mu\nu}^a(A)$ [3, 5].  However, the propagator for the quantum field $Q_\mu^a$ can develop an effective mass if $E^{ab} \neq 0$ in the long distance limit, so that the emission of low energy monochromatic quanta could serve as an infrared regulator in Yang-Mills theory.  The infrared problem has recently been investigated in close detail [9].

\section*{Acknowledgements}
R. Macleod was helpful in the early stages of this work.

\end{document}